\documentclass[twocolumn,superscriptaddress,altaffilletter,prc]{revtex4-1}

\usepackage{graphicx}
\usepackage{amssymb}
\usepackage{amsmath}
\usepackage{url}

\usepackage{xcolor}

\usepackage[perpage]{footmisc}

\begin{document}


\title{Search for two neutrino double electron capture of $^{124}$Xe and $^{126}$Xe in the full exposure of the LUX detector}


\author{D.S.~Akerib} 
\affiliation{SLAC National Accelerator Laboratory, 2575 Sand Hill Road, Menlo Park, CA 94205, USA} 
\affiliation{Kavli Institute for Particle Astrophysics and Cosmology, Stanford University, 452 Lomita Mall, Stanford, CA 94309, USA} 
\author{S.~Alsum} \affiliation{University of Wisconsin-Madison, Department of Physics, 1150 University Ave., Madison, WI 53706, USA}  
\author{H.M.~Ara\'{u}jo} \affiliation{Imperial College London, High Energy Physics, Blackett Laboratory, London SW7 2BZ, United Kingdom}  
\author{X.~Bai} \affiliation{South Dakota School of Mines and Technology, 501 East St Joseph St., Rapid City, SD 57701, USA}  
\author{J.~Balajthy} \affiliation{University of California Davis, Department of Physics, One Shields Ave., Davis, CA 95616, USA}  
\author{A.~Baxter} \affiliation{University of Liverpool, Department of Physics, Liverpool L69 7ZE, UK}  
\author{E.P.~Bernard} \affiliation{University of California Berkeley, Department of Physics, Berkeley, CA 94720, USA}  
\author{A.~Bernstein} \affiliation{Lawrence Livermore National Laboratory, 7000 East Ave., Livermore, CA 94551, USA}  
\author{T.P.~Biesiadzinski} \affiliation{SLAC National Accelerator Laboratory, 2575 Sand Hill Road, Menlo Park, CA 94205, USA} \affiliation{Kavli Institute for Particle Astrophysics and Cosmology, Stanford University, 452 Lomita Mall, Stanford, CA 94309, USA} 
\author{E.M.~Boulton} \affiliation{University of California Berkeley, Department of Physics, Berkeley, CA 94720, USA} \affiliation{Lawrence Berkeley National Laboratory, 1 Cyclotron Rd., Berkeley, CA 94720, USA} \affiliation{Yale University, Department of Physics, 217 Prospect St., New Haven, CT 06511, USA}
\author{B.~Boxer} \affiliation{University of Liverpool, Department of Physics, Liverpool L69 7ZE, UK}  
\author{P.~Br\'as} \affiliation{LIP-Coimbra, Department of Physics, University of Coimbra, Rua Larga, 3004-516 Coimbra, Portugal}  
\author{S.~Burdin} \affiliation{University of Liverpool, Department of Physics, Liverpool L69 7ZE, UK}  
\author{D.~Byram} \affiliation{University of South Dakota, Department of Physics, 414E Clark St., Vermillion, SD 57069, USA} \affiliation{South Dakota Science and Technology Authority, Sanford Underground Research Facility, Lead, SD 57754, USA} 
\author{M.C.~Carmona-Benitez} \affiliation{Pennsylvania State University, Department of Physics, 104 Davey Lab, University Park, PA  16802-6300, USA}  
\author{C.~Chan} \affiliation{Brown University, Department of Physics, 182 Hope St., Providence, RI 02912, USA}  
\author{J.E.~Cutter} \affiliation{University of California Davis, Department of Physics, One Shields Ave., Davis, CA 95616, USA}  
\author{L.~de\,Viveiros}  \affiliation{Pennsylvania State University, Department of Physics, 104 Davey Lab, University Park, PA  16802-6300, USA}  
\author{E.~Druszkiewicz} \affiliation{University of Rochester, Department of Physics and Astronomy, Rochester, NY 14627, USA}  
\author{A.~Fan} \affiliation{SLAC National Accelerator Laboratory, 2575 Sand Hill Road, Menlo Park, CA 94205, USA} \affiliation{Kavli Institute for Particle Astrophysics and Cosmology, Stanford University, 452 Lomita Mall, Stanford, CA 94309, USA} 
\author{S.~Fiorucci} \affiliation{Lawrence Berkeley National Laboratory, 1 Cyclotron Rd., Berkeley, CA 94720, USA} \affiliation{Brown University, Department of Physics, 182 Hope St., Providence, RI 02912, USA} 
\author{R.J.~Gaitskell} \affiliation{Brown University, Department of Physics, 182 Hope St., Providence, RI 02912, USA}  
\author{C.~Ghag} \affiliation{Department of Physics and Astronomy, University College London, Gower Street, London WC1E 6BT, United Kingdom}  
\author{M.G.D.~Gilchriese} \affiliation{Lawrence Berkeley National Laboratory, 1 Cyclotron Rd., Berkeley, CA 94720, USA}  
\author{C.~Gwilliam} \affiliation{University of Liverpool, Department of Physics, Liverpool L69 7ZE, UK}  
\author{C.R.~Hall} \affiliation{University of Maryland, Department of Physics, College Park, MD 20742, USA}  
\author{S.J.~Haselschwardt} \affiliation{University of California Santa Barbara, Department of Physics, Santa Barbara, CA 93106, USA}  
\author{S.A.~Hertel} \affiliation{University of Massachusetts, Amherst Center for Fundamental Interactions and Department of Physics, Amherst, MA 01003-9337 USA} \affiliation{Lawrence Berkeley National Laboratory, 1 Cyclotron Rd., Berkeley, CA 94720, USA} 
\author{D.P.~Hogan} \affiliation{University of California Berkeley, Department of Physics, Berkeley, CA 94720, USA}  
\author{M.~Horn} \affiliation{South Dakota Science and Technology Authority, Sanford Underground Research Facility, Lead, SD 57754, USA} \affiliation{University of California Berkeley, Department of Physics, Berkeley, CA 94720, USA} 
\author{D.Q.~Huang} \affiliation{Brown University, Department of Physics, 182 Hope St., Providence, RI 02912, USA}  
\author{C.M.~Ignarra} \affiliation{SLAC National Accelerator Laboratory, 2575 Sand Hill Road, Menlo Park, CA 94205, USA} \affiliation{Kavli Institute for Particle Astrophysics and Cosmology, Stanford University, 452 Lomita Mall, Stanford, CA 94309, USA} 
\author{R.G.~Jacobsen} \affiliation{University of California Berkeley, Department of Physics, Berkeley, CA 94720, USA}  
\author{O.~Jahangir} \affiliation{Department of Physics and Astronomy, University College London, Gower Street, London WC1E 6BT, United Kingdom}  
\author{W.~Ji} \affiliation{SLAC National Accelerator Laboratory, 2575 Sand Hill Road, Menlo Park, CA 94205, USA} \affiliation{Kavli Institute for Particle Astrophysics and Cosmology, Stanford University, 452 Lomita Mall, Stanford, CA 94309, USA} 
\author{K.~Kamdin} \affiliation{University of California Berkeley, Department of Physics, Berkeley, CA 94720, USA} \affiliation{Lawrence Berkeley National Laboratory, 1 Cyclotron Rd., Berkeley, CA 94720, USA} 
\author{K.~Kazkaz} \affiliation{Lawrence Livermore National Laboratory, 7000 East Ave., Livermore, CA 94551, USA}  
\author{D.~Khaitan} \affiliation{University of Rochester, Department of Physics and Astronomy, Rochester, NY 14627, USA}  
\author{E.V.~Korolkova} \affiliation{University of Sheffield, Department of Physics and Astronomy, Sheffield, S3 7RH, United Kingdom}  
\author{S.~Kravitz} \affiliation{Lawrence Berkeley National Laboratory, 1 Cyclotron Rd., Berkeley, CA 94720, USA}  
\author{V.A.~Kudryavtsev} \affiliation{University of Sheffield, Department of Physics and Astronomy, Sheffield, S3 7RH, United Kingdom}  
\author{E.~Leason} \affiliation{SUPA, School of Physics and Astronomy, University of Edinburgh, Edinburgh EH9 3FD, United Kingdom}  
\author{B.G.~Lenardo} \affiliation{University of California Davis, Department of Physics, One Shields Ave., Davis, CA 95616, USA} \affiliation{Lawrence Livermore National Laboratory, 7000 East Ave., Livermore, CA 94551, USA} 
\author{K.T.~Lesko} \affiliation{Lawrence Berkeley National Laboratory, 1 Cyclotron Rd., Berkeley, CA 94720, USA}  
\author{J.~Liao} \affiliation{Brown University, Department of Physics, 182 Hope St., Providence, RI 02912, USA}  
\author{J.~Lin} \affiliation{University of California Berkeley, Department of Physics, Berkeley, CA 94720, USA}  
\author{A.~Lindote} 
\email[Corresponding author: ] {alex@coimbra.lip.pt}
\affiliation{LIP-Coimbra, Department of Physics, University of Coimbra, Rua Larga, 3004-516 Coimbra, Portugal}  
\author{M.I.~Lopes} \affiliation{LIP-Coimbra, Department of Physics, University of Coimbra, Rua Larga, 3004-516 Coimbra, Portugal}  
\author{A.~Manalaysay} \affiliation{University of California Davis, Department of Physics, One Shields Ave., Davis, CA 95616, USA}  
\author{R.L.~Mannino} \affiliation{Texas A \& M University, Department of Physics, College Station, TX 77843, USA} \affiliation{University of Wisconsin-Madison, Department of Physics, 1150 University Ave., Madison, WI 53706, USA} 
\author{N.~Marangou} \affiliation{Imperial College London, High Energy Physics, Blackett Laboratory, London SW7 2BZ, United Kingdom}  
\author{M.F.~Marzioni} \affiliation{SUPA, School of Physics and Astronomy, University of Edinburgh, Edinburgh EH9 3FD, United Kingdom}  
\author{D.N.~McKinsey} \affiliation{University of California Berkeley, Department of Physics, Berkeley, CA 94720, USA} \affiliation{Lawrence Berkeley National Laboratory, 1 Cyclotron Rd., Berkeley, CA 94720, USA} 
\author{D.-M.~Mei} \affiliation{University of South Dakota, Department of Physics, 414E Clark St., Vermillion, SD 57069, USA}  
\author{M.~Moongweluwan} \affiliation{University of Rochester, Department of Physics and Astronomy, Rochester, NY 14627, USA}  
\author{J.A.~Morad} \affiliation{University of California Davis, Department of Physics, One Shields Ave., Davis, CA 95616, USA}  
\author{A.St.J.~Murphy} 
\email[Corresponding author: ] {a.s.murphy@ed.ac.uk}
\affiliation{SUPA, School of Physics and Astronomy, University of Edinburgh, Edinburgh EH9 3FD, United Kingdom}  
\author{A.~Naylor} \affiliation{University of Sheffield, Department of Physics and Astronomy, Sheffield, S3 7RH, United Kingdom}  
\author{C.~Nehrkorn} \affiliation{University of California Santa Barbara, Department of Physics, Santa Barbara, CA 93106, USA}  
\author{H.N.~Nelson} \affiliation{University of California Santa Barbara, Department of Physics, Santa Barbara, CA 93106, USA}  
\author{F.~Neves} \affiliation{LIP-Coimbra, Department of Physics, University of Coimbra, Rua Larga, 3004-516 Coimbra, Portugal}  
\author{A.~Nilima} \affiliation{SUPA, School of Physics and Astronomy, University of Edinburgh, Edinburgh EH9 3FD, United Kingdom}  
\author{K.C.~Oliver-Mallory} \affiliation{University of California Berkeley, Department of Physics, Berkeley, CA 94720, USA} \affiliation{Lawrence Berkeley National Laboratory, 1 Cyclotron Rd., Berkeley, CA 94720, USA} 
\author{K.J.~Palladino} \affiliation{University of Wisconsin-Madison, Department of Physics, 1150 University Ave., Madison, WI 53706, USA}  
\author{E.K.~Pease} \affiliation{University of California Berkeley, Department of Physics, Berkeley, CA 94720, USA} \affiliation{Lawrence Berkeley National Laboratory, 1 Cyclotron Rd., Berkeley, CA 94720, USA} 
\author{Q.~Riffard} \affiliation{University of California Berkeley, Department of Physics, Berkeley, CA 94720, USA} \affiliation{Lawrence Berkeley National Laboratory, 1 Cyclotron Rd., Berkeley, CA 94720, USA} 
\author{G.R.C.~Rischbieter} \affiliation{University at Albany, State University of New York, Department of Physics, 1400 Washington Ave., Albany, NY 12222, USA}  
\author{C.~Rhyne} \affiliation{Brown University, Department of Physics, 182 Hope St., Providence, RI 02912, USA}  
\author{P.~Rossiter} \affiliation{University of Sheffield, Department of Physics and Astronomy, Sheffield, S3 7RH, United Kingdom}  
\author{S.~Shaw} \affiliation{University of California Santa Barbara, Department of Physics, Santa Barbara, CA 93106, USA} \affiliation{Department of Physics and Astronomy, University College London, Gower Street, London WC1E 6BT, United Kingdom} 
\author{T.A.~Shutt} \affiliation{SLAC National Accelerator Laboratory, 2575 Sand Hill Road, Menlo Park, CA 94205, USA} \affiliation{Kavli Institute for Particle Astrophysics and Cosmology, Stanford University, 452 Lomita Mall, Stanford, CA 94309, USA} 
\author{C.~Silva} \affiliation{LIP-Coimbra, Department of Physics, University of Coimbra, Rua Larga, 3004-516 Coimbra, Portugal}  
\author{M.~Solmaz} \affiliation{University of California Santa Barbara, Department of Physics, Santa Barbara, CA 93106, USA}  
\author{V.N.~Solovov} \affiliation{LIP-Coimbra, Department of Physics, University of Coimbra, Rua Larga, 3004-516 Coimbra, Portugal}  
\author{P.~Sorensen} \affiliation{Lawrence Berkeley National Laboratory, 1 Cyclotron Rd., Berkeley, CA 94720, USA}  
\author{T.J.~Sumner} \affiliation{Imperial College London, High Energy Physics, Blackett Laboratory, London SW7 2BZ, United Kingdom}  
\author{M.~Szydagis} \affiliation{University at Albany, State University of New York, Department of Physics, 1400 Washington Ave., Albany, NY 12222, USA}  
\author{D.J.~Taylor} \affiliation{South Dakota Science and Technology Authority, Sanford Underground Research Facility, Lead, SD 57754, USA}  
\author{R.~Taylor} \affiliation{Imperial College London, High Energy Physics, Blackett Laboratory, London SW7 2BZ, United Kingdom}  
\author{W.C.~Taylor} \affiliation{Brown University, Department of Physics, 182 Hope St., Providence, RI 02912, USA}  
\author{B.P.~Tennyson} \affiliation{Yale University, Department of Physics, 217 Prospect St., New Haven, CT 06511, USA}  
\author{P.A.~Terman} \affiliation{Texas A \& M University, Department of Physics, College Station, TX 77843, USA}  
\author{D.R.~Tiedt} \affiliation{University of Maryland, Department of Physics, College Park, MD 20742, USA}  
\author{W.H.~To} \affiliation{California State University Stanislaus, Department of Physics, 1 University Circle, Turlock, CA 95382, USA}  
\author{M.~Tripathi} \affiliation{University of California Davis, Department of Physics, One Shields Ave., Davis, CA 95616, USA}  
\author{L.~Tvrznikova} \affiliation{University of California Berkeley, Department of Physics, Berkeley, CA 94720, USA} \affiliation{Lawrence Berkeley National Laboratory, 1 Cyclotron Rd., Berkeley, CA 94720, USA} \affiliation{Yale University, Department of Physics, 217 Prospect St., New Haven, CT 06511, USA}
\author{U.~Utku} \affiliation{Department of Physics and Astronomy, University College London, Gower Street, London WC1E 6BT, United Kingdom}  
\author{S.~Uvarov} \affiliation{University of California Davis, Department of Physics, One Shields Ave., Davis, CA 95616, USA}  
\author{A.~Vacheret} \affiliation{Imperial College London, High Energy Physics, Blackett Laboratory, London SW7 2BZ, United Kingdom}  
\author{V.~Velan} \affiliation{University of California Berkeley, Department of Physics, Berkeley, CA 94720, USA}  
\author{R.C.~Webb} \affiliation{Texas A \& M University, Department of Physics, College Station, TX 77843, USA}  
\author{J.T.~White} \affiliation{Texas A \& M University, Department of Physics, College Station, TX 77843, USA}  
\author{T.J.~Whitis} \affiliation{SLAC National Accelerator Laboratory, 2575 Sand Hill Road, Menlo Park, CA 94205, USA} \affiliation{Kavli Institute for Particle Astrophysics and Cosmology, Stanford University, 452 Lomita Mall, Stanford, CA 94309, USA} 
\author{M.S.~Witherell} \affiliation{Lawrence Berkeley National Laboratory, 1 Cyclotron Rd., Berkeley, CA 94720, USA}  
\author{F.L.H.~Wolfs} \affiliation{University of Rochester, Department of Physics and Astronomy, Rochester, NY 14627, USA}  
\author{D.~Woodward} \affiliation{Pennsylvania State University, Department of Physics, 104 Davey Lab, University Park, PA  16802-6300, USA}  
\author{J.~Xu} \affiliation{Lawrence Livermore National Laboratory, 7000 East Ave., Livermore, CA 94551, USA}  
\author{C.~Zhang} \affiliation{University of South Dakota, Department of Physics, 414E Clark St., Vermillion, SD 57069, USA}  

\collaboration{The LUX Collaboration}

\date{\today}

\begin{abstract}
Two-neutrino double electron capture is a process allowed in the Standard Model of Particle Physics. This rare decay has been observed in $^{78}$Kr, $^{130}$Ba and more recently in $^{124}$Xe. In this publication we report on the search for this process in $^{124}$Xe and $^{126}$Xe using the full exposure of the Large Underground Xenon (LUX) experiment, in a total of of 27769.5~kg-days. No evidence of a signal was observed, allowing us to set 90\% C.L. lower limits for the half-lives of these decays of $2.0\times10^{21}$~years for $^{124}$Xe and $1.9\times10^{21}$~years for $^{126}$Xe.
\end{abstract}

\maketitle


\section{Introduction}
\label{Sec:Intro}
\noindent An undiscovered property of the neutrino is its Majorana nature - is it its own antiparticle~\cite{jr:pontecorvo, jr:DBD, jr:neutrinolessDBD, jr:neutrinolessDBD_2}?
An equivalent question is whether the mass of the neutrino is purely of Dirac type, or whether there is also a Majorana mass contribution. If the latter, then neutrinos may participate in lepton-number violating processes, with consequences including their possible seeding of a matter-antimatter asymmetry in the extreme environment of the early universe~\cite{jr:luty}. 
Experimental access to this problem is mainly being sought through searches for neutrinoless double beta decay (0$\nu\beta\beta$)~\cite{jr:DBD, jr:neutrinolessDBD, jr:neutrinolessDBD_2}. 
Neutrinoless double electron capture (0$\nu$DEC) is a second-order weak nuclear process that, if observed, would similarly indicate a Majorana mass contribution~\cite{jr:Sujkowski}.
In most situations the lifetime for this process is expected to be several orders of magnitude larger than that of 0$\nu\beta\beta$, but a possible resonant 0$\nu$DEC process might occur~\cite{jr:neutrinolessDEC} in which the close degeneracy of the initial and final (excited) atomic states could enhance the decay rate by a factor as large as 10$^{6}$, making it observable.  
A number of studies have since pursued this possibility~\cite{jr:Eliseev2012, jr:Krivoruchenko2011, jr:Sujkowski, jr:neutrinolessDEC, jr:Suhonen_3, jr:Eliseev, jr:Green, jr:Barabash, jr:Barabash_2, jr:Kidd}.

Simultaneous capture of two electrons with the emission of two neutrinos, {\it i.e.} two-neutrino double electron capture (2$\nu$DEC), is a related decay process that is allowed by the standard model~\cite{jr:Meyerhof, jr:Warczak, jr:Yakhontov}. 
This decay mode has been observed in $^{78}$Kr~\cite{jr:Gavrilyuk}, $^{130}$Ba~\cite{jr:Meshik, jr:Pujol} and more recently in $^{124}$Xe~\cite{jr:Fieguth} with measured lifetimes in the range of (0.6 -- 18) $\times$10$^{21}$ years. 
These lifetimes provide knowledge of the underlying nuclear matrix element, $M_{2\nu}$, through the relationship

\begin{equation}
(T_{\frac{1}{2},2\nu})^{-1}=G^{2EC}_{2\nu}{g^4_{A}} | m_{e} c^2 M_{2\nu} |^2,
\label{eq:matrixelement2}
\end{equation}

\noindent where $g_{A}$ is the axial coupling constant and $G^{2EC}_{2\nu}$ is the phase space factor for this decay \cite{jr:kotila2013}.
Comparison of experimentally informed nuclear matrix elements of two neutrino double beta decay or double electron capture with the predictions of nuclear models provides a valuable test for these models, which may then use this information to improve their calculations of the unknown nuclear matrix elements of neutrinoless double-beta decay~\cite{jr:Bilenky, jr:Simkovic1, jr:Simkovic2, jr:Fieguth}. Searches for 2$\nu$DEC provide particular input to the calculation of nuclear matrix elements on the proton-rich side of the mass parabola for even--even isobars~\cite{jr:Suhonen}. 

Natural xenon consists of eight isotopes, including 0.095\% $^{124}$Xe and 0.089\% $^{126}$Xe (by mass) that may undergo 2$\nu$DEC. The decay process is similar for both isotopes:
\begin{equation}
^{A}Xe + 2e^{-} \rightarrow~^{A}Te + 2\nu_{e},
\label{eq:xe124}
\end{equation}
\noindent where A is the atomic number. The reaction Q-value is 2864 keV for $^{124}$Xe and 919~keV for $^{126}$Xe. This energy is carried by the neutrinos, which easily escape without being detected, while the nuclear recoil energy is of the order of 30~eV and therefore negligible. The atomic de-excitation following the double electron capture leads to a cascade of X-rays and Auger electrons being emitted from the daughter Te atom as the newly created vacancies are refilled. The branching ratio for events in which both captured electrons are from the K-shell dominates, and leads to a total energy deposit signal of 64.57~keV of which 64.3~keV can be detected \cite{jr:Fieguth} (the difference corresponding to energy depositions at the end of the cascade that are too small to produce ionisation or scintillation in the xenon). The signal region will be dominated by $^{124}$Xe, due to the $Q^5$ dependence of the phase space factor.

In fact, $^{124}$Xe has the highest Q-value amongst DEC candidate nuclei, opening the possibility for two additional decay channels: $2\beta^{+}$ (possible in only six nuclides) and the mixed mode $\beta^{+} EC$. Since the nuclear matrix elements for each of the three decay modes will be similar~\cite{jr:Pirinen2015}, the lifetimes may be expected to differ only by factors determined by phase space differences~\cite{jr:kotila2013}. Consequently, one might expect lifetimes of the order of $10^{23}$~years for $\beta^{+} EC$ and $10^{27}$~years for $2\beta^{+}$, and their detection is therefore out of the reach of LUX. On the other hand, they have very distinct topologies, with multiple energy depositions per decay (from the atomic de-excitation, the kinetic energy of the positrons and the 511~keV $\gamma$-rays from the annihilations), which can be explored to discriminate them from the background with high efficiency in larger detectors.

Searches for 2$\nu$DEC  in Xe typically look for the mode with both captured electrons coming from the K-shell (2$\nu KK$). Several of these searches reported no indications of a signal and set lower limits for its half-life (\textit{e.g.} \cite{jr:XMASS2018, jr:XENON100, jr:Gavriljuk2018}), with XMASS achieving the most stringent (90\% C.L.) limits at 2.1$\times$10$^{22}$~years for $^{124}$Xe and 1.9$\times$10$^{22}$~years for $^{126}$Xe. Recently XENON1T, which uses the same detector technology as LUX but has a larger fiducial mass of 1500~kg, announced the first observation of the 2$\nu KK$ mode of this decay in $^{124}$Xe with a half-life of $1.8\times10^{22}$~years and a 4.4$\sigma$ statistical significance~\cite{jr:Fieguth}. 

In this work we examine data from the Large Underground Xenon detector (LUX) for evidence of 2$\nu KK$ decay from $^{124}$Xe or $^{126}$Xe. Details of the experimental approach, of the LUX detector and the specific conditions in which the data used in this analysis was acquired are discussed in Section~\ref{Sec:Exp}, while the details of the data analysis and the statistical method used are presented in Section~\ref{Sec:Analysis}. Finally, results from this study are presented and discussed in Section~\ref{Sec:Results}.

\section{Experiment Details}
\label{Sec:Exp}

\noindent The detection of the atomic de-excitation signal following 2$\nu KK$ decay is achievable in large-scale two-phase time projection chambers (TPCs) used as dark matter detectors such as LUX, as was first proposed by Mei {\it et al.}~\cite{jr:Mei} and Barros {\it et al.}~\cite{jr:Barros} and recently confirmed by the XENON1T collaboration~\cite{jr:Fieguth}. These detectors provide excellent sensitivity to energy depositions of this magnitude and, despite the relatively low isotopic abundance of both isotopes, have targets of sufficient size to provide a significant isotope-specific exposure.

The LUX detector has been extensively described in~\cite{jr:LUXdet2013}. Briefly, it consists of a low-radioactivity titanium vessel partially filled with liquid xenon such that above the liquid a thin layer of gaseous xenon is maintained. A vertical electric field of 181~V/cm is established in the xenon target volume (with a mass of 250~kg of natural xenon) via a gate grid placed just below the liquid surface and a cathode at the base of the liquid (total drift length of 48.3~cm). 
Energy depositions in the xenon target lead to scintillation, ionization and heat. The scintillation, which has a narrow band of wavelengths centered at 175~nm, is directly observed 
by UV-sensitive high quantum efficiency photomultiplier tubes (PMTs) arranged in two arrays above and below the xenon volume (producing what is usually referred to as the S1 signal). Simultaneously, the electric field drifts a fraction of the ionized electrons to the liquid surface, where a stronger field extracts them to the gaseous xenon region where they produce secondary electroluminescence that is also registered by the PMTs (S2 signal). Horizontal positions are reconstructed using the S2 light distribution in the top array~\cite{jr:mercury}, while the delay (0~--~322~$\mathrm{\mu}$s) between the S1 and the S2 provides the vertical position~\cite{jr:LUXresol}.
Single electrons extracted from the liquid produce signals with an average of 24.7 detected photons~\cite{jr:PRD} and are easily observed, providing a very low energy threshold for experimental searches (see \textit{e.g.}~\cite{jr:Xe127_LUX}). The instrument is immersed in a 7.6~m diameter and 6.1~m high tank with ultrapure water to reduce the background from external sources, and installed at a depth of 4850~feet at the Sanford Underground Research Facility, the underground location providing shielding equivalent to 4300~m of water from cosmic radiation (a reduction of $\mathcal{O}(10^{-7})$ in the rate of cosmic muons compared to the surface)~\cite{jr:SURFmuons}.

LUX was operated from 2013 through to 2016, in two separate WIMP search campaigns. The first, between April and August 2013 (WS2013), collected 95 live-days of data, while a second longer run (WS2014-16) between September 2014 and May 2016 collected 332 live-days.
As a result of an \textit{in-situ} grid ``conditioning" performed before the start of the second run significant changes were observed in the paths of the drifting electrons, from having been near-vertical in WS2013, to being strongly deflected by electric charge of the S2 signals building up on the walls of the detector during WS2014-16~\cite{jr:combinedLUX, jr:LUXfield}.
The accumulated charge leads to a time and space varying electric field magnitude, with direct impact in the recombination of electron-ion pairs and thus the light and charge yields. Section \ref{Sec:Analysis} details how these effects were dealt with in data analysis.

\section{Data Analysis}
\label{Sec:Analysis}

\subsection{Data and event selection}
\label{SSec:DataSelection}
\noindent Events included in this analysis were selected by requiring that the PMT waveforms contain a single S1 and a single S2, as the timescale for the atomic de-excitation following 2$\nu KK$ decay and the range of the emitted X-rays and Auger electrons are too short for individual interactions to be resolved in the detector. Safeguards were put in place to prevent events in which the main S1 and/or S2 are followed by small spurious pulses from being discarded. More specifically, except at very small energy depositions ($\lesssim$10~keV), S2 pulses are typically followed by a trail of single electrons extracted from the liquid, which are the result of photoionisation caused by the S2 light in the grids and in impurities in the xenon bulk. These isolated single electrons can overlap in time and produce pulses that resemble legitimate S2s following the main S2, leading to the event being mistakenly identified as a multiple scatter and thus discarded. If not accounted for, this effect leads to an efficiency penalty that increases with event energy, reaching $\sim$20\% at the 2$\nu KK$ energy. Allowing for additional small spurious pulses after the main S1 and S2 recovers the 98.8\% efficiency determined for low energy single scatter events~\cite{jr:PRD}. 

$^{83m}$Kr calibrations were carried out frequently throughout the detector operation, for calibration and monitoring. $^{83m}$Kr has a half-life of 1.83~h and decays via two transitions, of 32.1 and 9.4~keV, in quick succession (with an intervening half-life of 154~ns)~\cite{jr:Kr83_LUX}, which are often seen as a single interaction in the detector, with a total energy of 41.5~keV.
While such signals have been shown to not contaminate other rare event searches conducted by LUX, the summed energies of these decays, smeared by the experimental energy resolution, results in possible contamination of the control region used to estimate the background in the 2$\nu$KK region (which starts at 48.1~keV, as discussed in detail in Section~\ref{SSec:Rolke}). Consequently, the data used for this work excludes acquisition periods following $^{83m}$Kr injections such that possible contamination of the data by emissions from this isotope is at a negligible level compared to the background rate, at a large (34\%) live-time penalty.

An important background for this search results from the calibrations of the response of the LUX detector to nuclear recoils, which were performed at the end of WS2013 and at various occasions during the WS2014-16 campaign using a deuterium-deuterium (DD) neutron generator~\cite{jr:DD}. These calibrations lead to the production of unstable isotopes via neutron activation (\textit{e.g.} $^{125}$Xe, $^{127}$Xe, $^{129m}$Xe, $^{131m}$Xe, $^{133}$Xe, $^{135}$Xe and $^{137}$Xe). Of these, the isotope of most concern to this analysis is $^{125}$I, resulting from the decay of $^{125}$Xe produced by neutron activation of $^{124}$Xe:

\begin{align}
^{124}Xe + n &\longrightarrow\ ^{125}Xe + \gamma \\
^{125}Xe &\xrightarrow[{16.9~h}]{EC}\ ^{125}I + \nu_e + \gamma + X
\end{align}

\noindent where X represents the set of X-rays and Auger electrons from the atomic de-excitation. $^{125}$I decays to the 35.5~keV excited state of $^{125}$Te via EC, with the resulting nuclear transition and atomic de-excitation leading to a total energy deposit of 67.3~keV in most decays (when the EC is from the K-shell, which occurs in $>$80\% of the decays)~\cite{site:nucleide_125I}:
\begin{align}
^{125}I &\xrightarrow[{59.4~d}]{EC}\ ^{125}Te  + \nu_e + \gamma + X\qquad(67.3~keV)
\end{align}

\noindent which in LUX is just slightly more than one sigma (2.7~keV) away from the 2$\nu$KK signal and therefore contaminates the signal region.
Furthermore, while $^{125}$Xe decays quickly with a half-life of only 16.9~h, $^{125}$I has a much longer half-life of 59.4~d --- but has been observed to be removed from LUX by the purification system (which uses a hot zirconium getter) with a decay constant of 5.4$\pm$0.4~d. Data with high $^{125}$I rate following DD calibration campaigns were used to characterise the energy resolution of the detector in this energy range (see Figure~\ref{fig:Eres} and discussion in Section~\ref{SSec:Energy}), but excluded from the 2$\nu$KK analysis in order to minimise contamination of the signal region. These exclusion periods were defined as 4 effective $^{125}$I removal periods (22 calendar days) after the end of DD calibrations.
Together with the exclusion periods following $^{83m}$Kr calibrations, this resulted in a final live-time of 242.2 days ($<$60\% of the available exposure).

Although neutron activation of Xe isotopes can also occur due to ambient neutrons in the underground laboratory~\cite{jr:neutron_flux} while xenon is outside the water tank during circulation, the small mass of this unshielded xenon and the reduced abundance of $^{124}$Xe in natural xenon make this mechanism for production of $^{125}$Xe negligible ($<$2 atoms per month) compared to the overall background.

\subsection{Energy calibration and resolution}
\label{SSec:Energy}

\begin{figure}[]
\centering
    \includegraphics[width=0.5\textwidth]{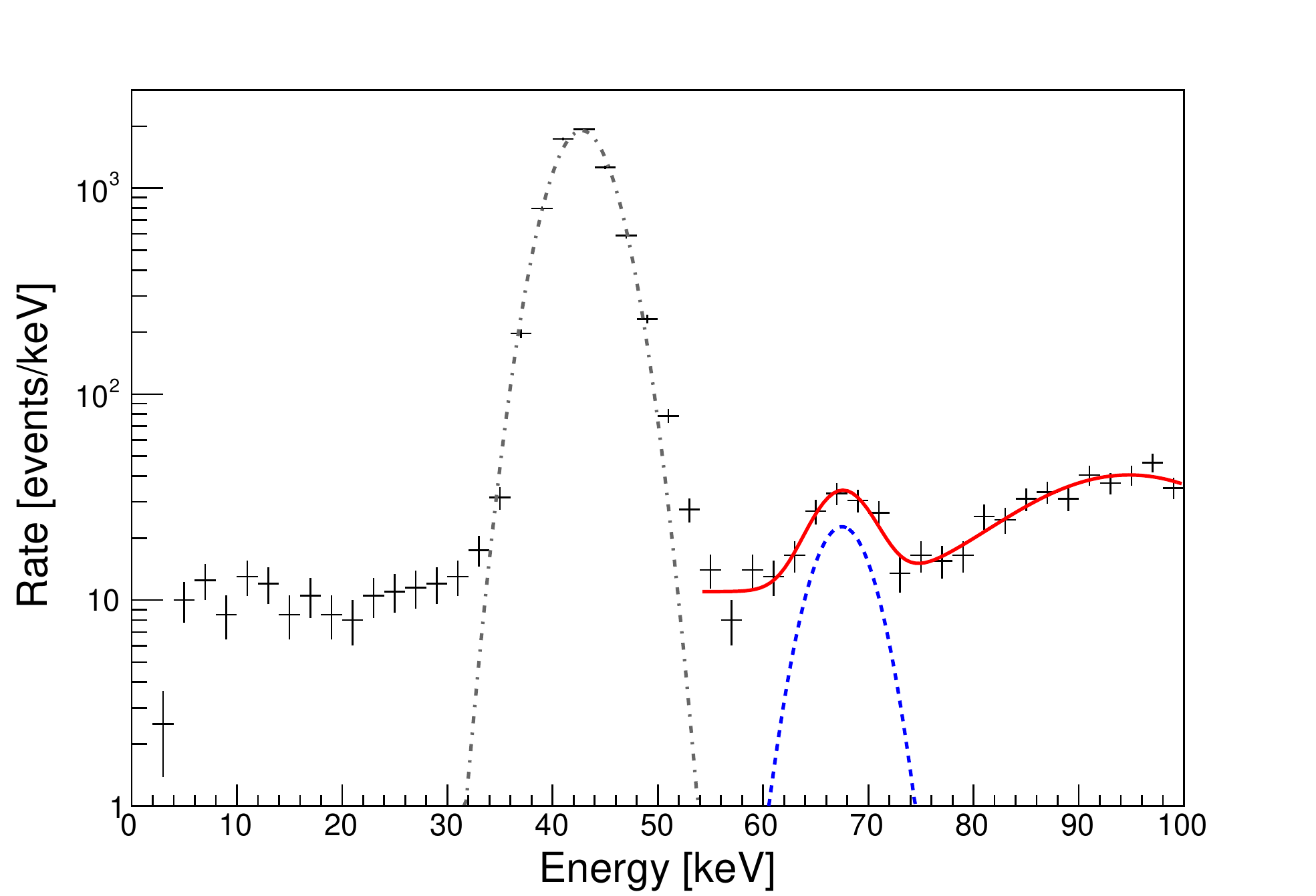}
    \caption{Energy spectrum obtained using data from two periods following nuclear recoil calibrations of the detector using the DD generator. A fit is done above 55~keV (shown by the continuous red curve) and includes a Gaussian to model the $^{125}$I decay peak (with a resolution of 4.2~\%, also shown by the blue dashed curve), a flat background component and an additional Gaussian to describe the $^{133}$Xe $\gamma$ (81~keV) + $\beta$ (346.4~keV end-point) decay also visible in these data. Also shown is a separate Gaussian fit to the $^{83m}$Kr region only (dashed-dotted line).}
    \label{fig:Eres}
\end{figure}

\noindent The energy has been reconstructed from S1 and S2 signal areas using the expression 

\begin{equation}
E = (S1_{c}/g_{1} + S2_{c}/(g_{2})) W,
\end{equation}

\noindent where $S1_{c}$ and $S2_{c}$ (measured in $phd$, representing the number of photons detected) are the S1 and the S2 signals corrected to equalize the detector response throughout the active volume; $g_{1}$ and $g_{2}$ are the gain factors, defined by the expectation values $\left\langle S1 \right\rangle = g_{1} n_{\gamma}$ and $\left\langle S2 \right\rangle = g_{2} n_{e}$, with $n_{\gamma}$ and $n_{e}$ representing the initial number of photons and electrons produced by the interaction, respectively; and $W = (13.7 \pm 0.2)$~eV is the mean work function for the production of either a photon or an electron in xenon (see~\cite{jr:NEST1} for a detailed discussion on this energy model). The energy calibration was based on fits to $\gamma$-ray lines in the 5.2 to 661.7~keV range, from decays of $^{127}$Xe, $^{83m}$Kr, $^{131m}$Xe, $^{129m}$Xe, $^{208}$Tl, $^{214}$Bi and $^{137}$Cs. The $g_{1}$ and $g_{2}$ values used for the WS2013 data are those reported in~\cite{jr:reanaLUX}, while in the WS2014-16 dataset they were found to slightly vary throughout the acquisition period~\cite{jr:combinedLUX}, with $g_{1}$ decreasing from 0.100$\pm$0.001 to 0.098$\pm$0.001 phd per emitted photon and $g_{2}$ from 19.40$\pm$0.45 to 18.75$\pm$0.29 phd per electron.
The periods following nuclear recoil calibrations of the detector with the DD generator have a high rate of $^{125}$I, offering an opportunity to estimate the energy resolution in this energy range. This was done using data from two high $^{125}$I rate periods, as shown in Figure~\ref{fig:Eres}, with a centroid of 67.5$\pm$0.4~keV and a resolution of 4.2$\pm$0.6\% (in good agreement with the expectation from~\cite{jr:LUXresol} for this energy). 

\subsection{Data Quality Cuts}
\label{SSec:DataCuts}

\noindent Events that passed the single scatter criterion were subjected to further data quality cuts, designed to remove pathological populations while minimising the impact in the signal detection efficiency. 
\begin{itemize}
\item We require a minimum size of 10~phd and 1000~phd for S1 and S2 pulse areas respectively, which removes random coincidence background but doesn’t impact the signal acceptance.
\item Energy depositions in the xenon gas produce a long continuous signal which can be classified as an S2 and mimic a single scatter event when accidentally paired with an isolated S1. This small population of events ($<$0.5\% in the region of interest) was excluded by requiring that $\sigma_{S2}>0.4~\mu s$ (where $\sigma_{S2}$ is the width of a Gaussian fit to the S2 pulse).
\item Finally, a cut was designed to remove a special class of multiple scatter events (dubbed gamma--X events) in which one of the interactions occurs in the active region of the detector (producing an S1 and an S2) and one or more energy depositions take place under the cathode, in the reverse field region, where only the S1 is detected. The multiple S1 pulses are merged and only one S2 is detected, leading to an incorrect energy reconstruction for the event.
\end{itemize}
Although mostly negligible in the low energy WIMP search range, the rate of these gamma-X events increases considerably for higher energies and is a relevant background in the 2$\nu$KK region. Occurring under the cathode, very close to the bottom PMT array, these additional scatters are expected to lead to events with a higher fraction of the S1 signal in a single bottom PMT compared to single scatters in the fiducial volume (well above the cathode), a feature that can be explored to identify them with high efficiency. 

\begin{figure}[]
\centering
    \includegraphics[width=0.5\textwidth]{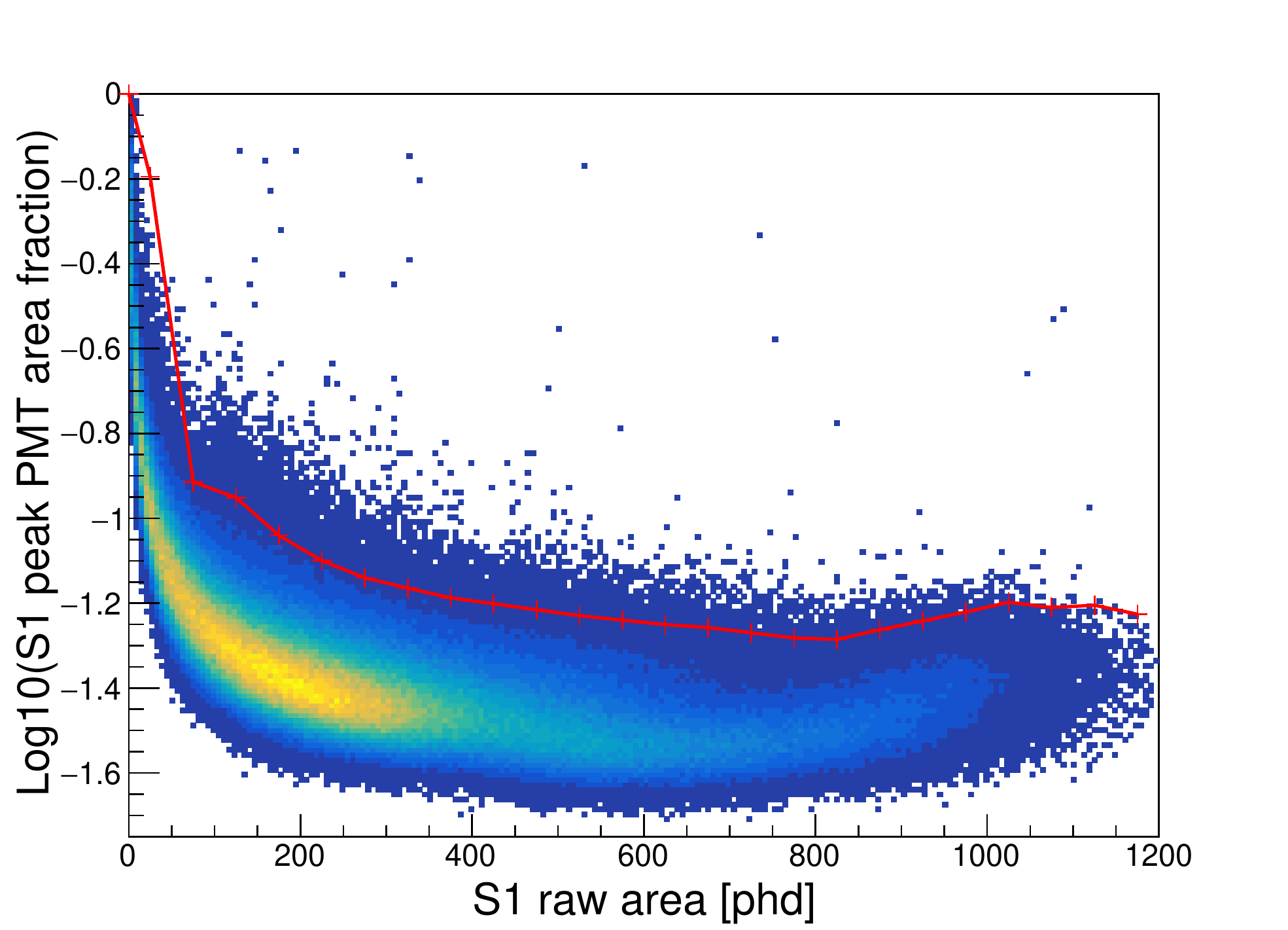}
    \includegraphics[width=0.5\textwidth]{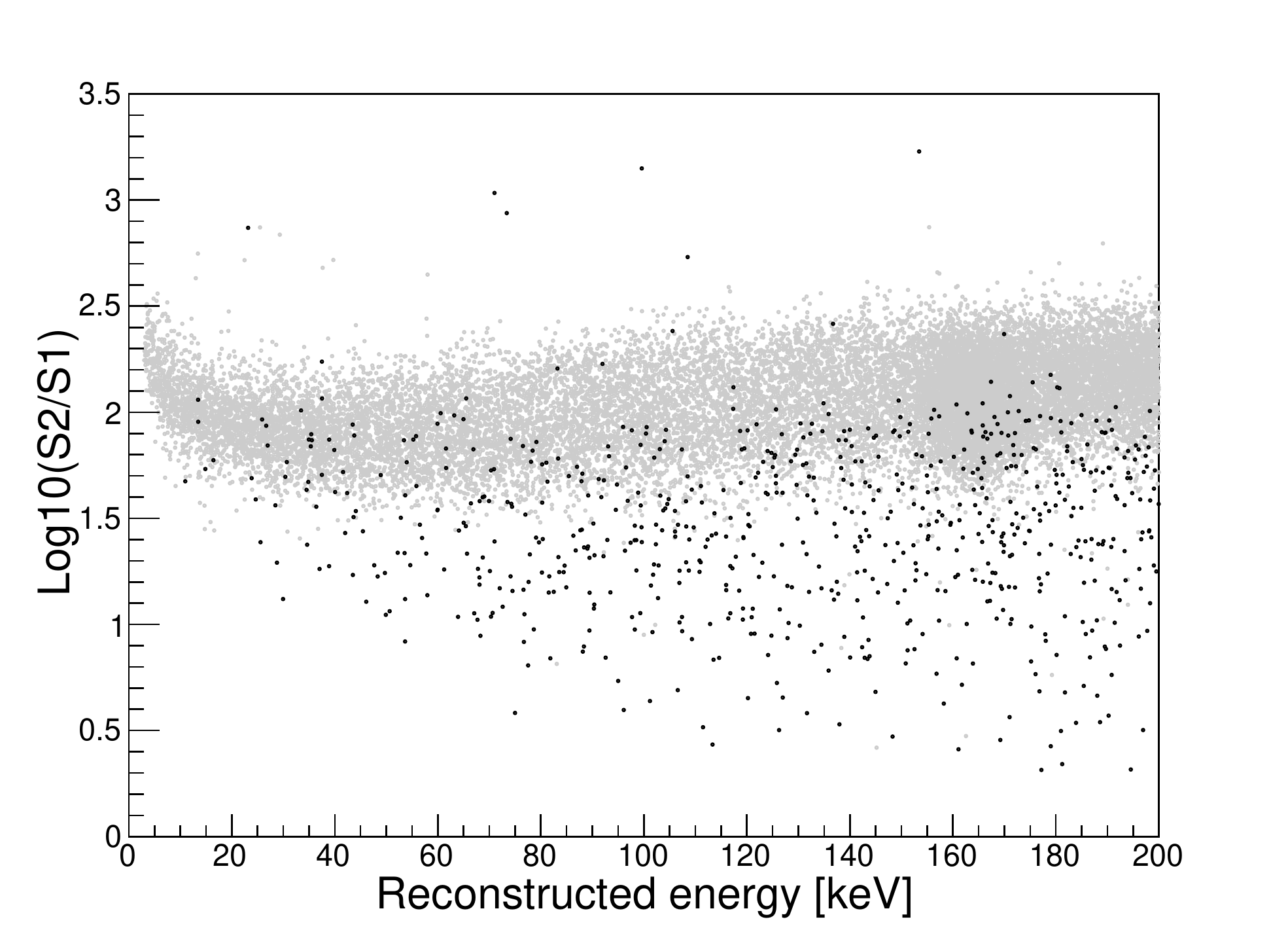}
    \caption{Definition of the Gamma-X cut and its effect in the LUX WS data. \textit{Top:} the cut to remove gamma-X events (shown by the red line) is defined using $^{14}$C data to minimise gamma--X contamination. It uses the highest fraction of the S1 raw area in a single bottom array PMT and is a function of the raw S1 area: the area fraction that keeps $>$99\% of the $^{14}$C population with a drift time longer than 250~$\mu s$ was found for each 50~phd wide raw S1 bin (shown by the red crosses); the final cut is then constructed using linear interpolations between the centers of these bins. \textit{Bottom:} events excluded by this cut (represented by the black dots) in the WS datasets are mostly in the region under the ER band, as expected.}
    \label{fig:gammaX}
\end{figure}

Data from the $^{14}$C calibration performed after the WS2014-16 run~\cite{jr:jon2019} was used to define this cut, as it provides a uniformly distributed population of events with energy depositions up to the 156.5~keV $\beta$-decay endpoint with minimal contamination of gamma--X events. As shown in the top panel of Figure~\ref{fig:gammaX}, the S1 raw area fraction in the bottom array PMT with the largest contribution to the S1 signal was plotted and binned as a function of the total uncorrected S1 area. The cut was then defined to keep $>$99\% of the $^{14}$C events in each raw S1 bin. As this fraction naturally increases with depth, and to maximise the signal acceptance efficiency, only events with a drift time longer than 250~$\mu s$ were used to define the cut. This resulted in an overall efficiency of 99.7\% at all depths (flat in the energy region of interest), but actually only events below $\sim$270~$\mu s$ are excluded (corresponding to the bottom 7.6~cm and 3~cm of the fiducial volumes in the WS2013 and WS2014-16 datasets, respectively). 

The bottom panel in Figure~\ref{fig:gammaX} shows the events from the WS datasets which are excluded by this cut, mainly in the region under the electron recoil band as expected. Consistent with being gamma-X events created by radioactive decays in the bottom PMT array, their density decreases exponentially with the distance from the bottom of the detector and they are mainly located above the innermost PMTs.

\subsection{Statistical Approach}
\label{SSec:Rolke}
\noindent The background in the 2$\nu$KK decay signal region is expected to be approximately flat, dominated by Compton scattering of high energy $\gamma$ rays and $\beta$ decays from contaminants mixed in the xenon (as discussed in more detail in Section~\ref{Sec:Results}). We therefore use the frequentist statistical approach of Rolke~{\it et al.}~\cite{jr:Rolke}, wherein a region of interest (ROI) is defined that broadly covers the signal region, and sidebands with the same width are defined to either side. The number of background events in the ROI is estimated as the average of the absolute number of events in the two side bands.
The Rolke statistical method is then used to estimate the number of signal events corresponding to the 90\% C.L. upper limit given the actual observation in the ROI, as well as the number of signal events corresponding to the sensitivity of the experiment (\textit{i.e.} the average expected limit in case of no signal), which can be used to estimate the corresponding limit and sensitivity on the half-life of the KK decay mode using:

\begin{equation}
T^{2\nu KK}_{1/2} \geq \frac{ln(2) a N_{A}}{A} \frac{\epsilon M \Delta T}{\mu_{up}}
\label{eq:halfLife}
\end{equation}

\noindent where $\mu_{up}$ is the upper limit on the number of signal events, a is the isotope abundance in natural xenon (0.095~\% for $^{124}$Xe and 0.089~\% for $^{126}$Xe) and A the respective molar mass (123.9~g/mol and 125.9~g/mol, respectively~\cite{tr:atomweights}), $N_A = 6.022\times10^{23}$ is Avogadro's number, $\epsilon$ is the signal detection efficiency, $\Delta T$ is the duration of the exposure and $M$ the fiducial mass.

Here, we define the ROI with 95.4\% signal acceptance as $\pm2\sigma$ from the signal median of 64.3~keV, which translates to a width of 10.8~keV for a 4.2\% energy resolution (measured from the fit to the $^{125}$I decay peak, see Figure~\ref{fig:Eres}), extending from 58.9 to 69.7~keV. The acceptance is reduced to 94.0\% after applying the flat efficiencies from the data quality cuts and the single scatter selection. The uncertainty in the acceptance of the ROI is calculated as having independent contributions from the error in the reconstruction of the mean energy and the error on the width of the peak, resulting in a final ROI acceptance $\epsilon=94.0^{+2.2}_{-3.2}$~\%.

\subsection{Fiducial volume definition}
\label{SSec:FVolume}

\noindent The fiducial volume used in this analysis was chosen in order to maximise the experiment sensitivity obtained from the Rolke statistical method, which depends only on the number of events observed in the side bands. This provides an unbiased optimisation strategy which takes into account both the fiducial mass and the corresponding background level. 

Given the increasing radial field observed in the WS2014-16 dataset a cylinder in S2-space~\cite{jr:combinedLUX} was chosen to define the volume, with the distances to the cathode, gate and PTFE walls optimised independently in an iterative process. 

For each tested combination, the fiducial mass was estimated using the fraction of $^{83m}$Kr decays in the volume relatively to those in the full active volume (which contains 250.9$\pm$2.1~kg of liquid xenon~\cite{jr:PRD}). This process was performed independently for the WS2013 dataset and for each of the WS2014-16 time bins, resulting in an overall exposure weighted fiducial mass of M=114.6$\pm$0.5~kg (containing 108.9$\pm$0.5~g of $^{124}$Xe and 102.0$\pm$0.5~g of $^{126}$Xe).
\section{Results and Discussion}
\label{Sec:Results}

\begin{figure}[]
\centering
    \includegraphics[width=0.5\textwidth]{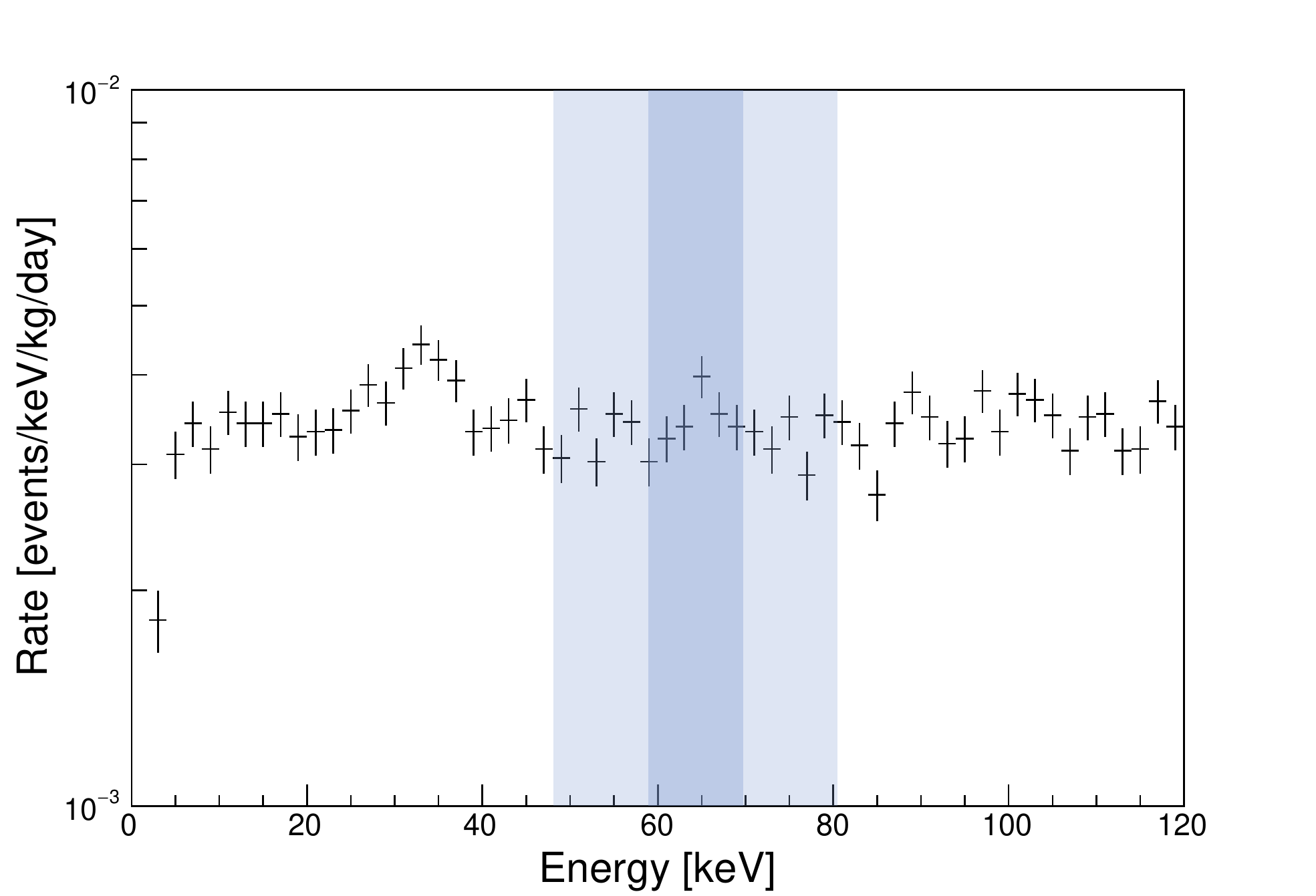}
    \caption{Event rate as a function of the deposition energy for the full 242.2~days LUX exposure. Only single scatter events outside the exclusion periods, passing all data quality cuts and within the fiducial volume defined for each time bin have been included. Also shown are the limits of the $\pm2\sigma$ ROI around the 2$\nu$KK signal (dark blue shaded region) and those of the two side bands (light blue shaded regions). Statistical (counting) errors are Poisson.}
    \label{fig:combined}
\end{figure}

\noindent Combining the data from the WS2013 and WS2014-16 datasets surviving the exclusion periods and data quality cuts produces the spectrum shown in Figure~\ref{fig:combined}. This is a broadly continuous background with a rate of 3.4$\times10^{-3}$ events/kg/keV/day, consistent with energy depositions from Compton scattering of high energy $\gamma$ rays from radiological (U, Th, $^{40}$K, $^{60}$Co) contamination in detector components and the ``naked" $\beta$-decay of $^{214}$Pb in the $^{222}$Rn chain and compatible with the expectation from the background model of the detector at low energies~\cite{jr:LUXbg}. The contribution from the $\beta$-decay of $^{85}$Kr is highly suppressed as a result of purification of the LUX xenon in a dedicated Kr-removal system prior to the detector installation underground~\cite{jr:LUXbg,jr:kr-removal}. A clear peak is seen around 32~keV due to the decay of the cosmogenically produced $^{127}$Xe ($T_\frac{1}{2}$=36.4~d) isotope, activated while the xenon was on the surface and therefore present at the start of the WS2013 dataset with an initial activity of 490~$\pm$~95~$\mu$Bq/kg~\cite{jr:LUXbg}.

A total of 993 and 991 events are observed in the left and right side bands, respectively, resulting in a background expectation of 992$\pm$31 events in the ROI and a 90\% C.L. sensitivity to the half-life of the $2\nu KK$ decay of $^{124}$Xe of $3.1 \times 10^{21}$~years (corresponding to an upper limit of 78 signal events). A small (1.3$\sigma$) upward fluctuation is visible in the ROI, with the 1031 events actually observed translating in a 90\% C.L. upper limit of 120 signal events and a corresponding lower limit for the half-life of this decay mode of $2.0 \times 10^{21}$~years. Similarly, the sensitivity to the half-life of $2\nu KK$ decay in $^{126}$Xe is $2.9 \times 10^{21}$~years while the limit is $1.9 \times 10^{21}$~years.

\section{Conclusions}
\label{Sec:Conclusions}

\noindent Data collected during the two science runs of the LUX detector were analysed to search for the $2\nu DEC$ decay of $^{124}$Xe and $^{126}$Xe, in particular the mode in which the two captured electrons are from the K-shell. No significant excess of events was observed in the 27769.5~kg-days total exposure, which allowed 90\% C.L. lower limits to be set for the half life of this decay mode of $2.0\times10^{21}$~years for $^{124}$Xe and $1.9\times10^{21}$~years for $^{126}$Xe. 

For the case of $^{124}$Xe this is an order of magnitude lower than the observed half-life reported by XENON1T, of $1.8\times10^{22}$~years (4.4~$\sigma$ significance) for a fiducial mass of 1.5~ton and 177.7~days exposure. This is a consequence of the much higher mass of XENON1T, which has a direct impact in the size of the available isotope samples but also allows for a more shielded fiducial volume and consequently a lower background rate. 

The next generation detector LUX-ZEPLIN~\cite{jr:LZWS}, with an active mass of 7~ton and an estimated fiducial mass in this energy region of 5.6~ton, will be able to confirm the observation of $2\nu KK$ decay of $^{124}$Xe from XENON1T and reach a discovery level 5$\sigma$-significance for this signal in just a few months of operation. Given its foreseen 1000 day long run it will also be able to search for the $2\nu \beta^{+}EC$ mixed decay mode ($10^{23}$~years), while the expected very long half life of the $2\nu \beta^{+}\beta^{+}$ channel ($10^{27}$~years) puts it out of reach of this generation of detectors. Similarly, the $2\nu DEC$ decay of $^{126}$Xe (with an estimated half-life of $\mathcal{O}(10^{24})$~years) will not lead to a statistically significant number of events overlapping the large $^{124}$Xe $2\nu DEC$ signal to claim an observation. 

 \section*{Acknowledgments}

The research supporting this work took place in whole or in part at the Sanford Underground Research Facility (SURF) in Lead, South Dakota. Funding for this work is supported by the U.S. Department of Energy, Office of Science, Office of High Energy Physics under Contract Numbers DE-SC0020216, DE-FG02-08ER41549, DE-FG02-91ER40688, DE-FG02-95ER40917, DE-FG02-91ER40674, DE-NA0000979, DE-FG02-11ER41738, DE-SC0015535, DE-SC0006605, DE-AC02-05CH11231, DE-AC52-07NA27344, DE-FG01-91ER40618 and DE-SC0019066. This research was also supported by the U.S. National Science Foundation under award numbers PHYS-0750671, PHY-0801536, PHY-1004661, PHY-1102470, PHY-1003660, PHY-1312561, PHY-1347449; the Research Corporation grant RA0350; the Center for Ultra-low Background Experiments in the Dakotas (CUBED); the South Dakota School of Mines and Technology (SDSMT). LIP-Coimbra acknowledges funding from Funda\c{c}\~{a}o para a Ci\^{e}ncia e a Tecnologia (FCT) through the project-grant PTDC/FIS-PAR/28567/2017. Imperial College and Brown University thank the UK Royal Society for travel funds under the International Exchange Scheme (IE120804). The UK groups acknowledge institutional support from Imperial College London, University College London and Edinburgh University, and from the Science \& Technology Facilities Council for PhD studentship ST/K502042/1 (AB). The University of Edinburgh is a charitable body, registered in Scotland, with registration number SC005336. The assistance of SURF and its personnel in providing physical access and general logistical and technical support is acknowledged.



\bibliographystyle{unsrt}
\bibliography{lux_2ec.bib}

\end{document}